\documentclass[11pt]{article}
\usepackage[nosort]{cite}
\usepackage{amsmath,amsfonts,amssymb}
\usepackage[small,bf,hang]{caption}
\usepackage{slashed}
\usepackage{mathabx}
\usepackage{latexsym,epsfig}
\usepackage{color}

\def\hybrid{
        \topmargin -20pt
        \oddsidemargin 0pt
        \headheight 0pt \headsep 0pt
        \textwidth 6.25in 
        \textheight 9.5in 
        \marginparwidth .875in
        \parskip 5pt plus 1pt \jot = 1.5ex}

\hybrid

 \csname
@addtoreset\endcsname{equation}{section}

\def\moth{\mathsurround=0pt}
\newdimen\zo \zo=0pt

\def\tick{\leaders\hrule height 0.5ex depth 0pt \hskip 0.5pt}
\def\upboxfill{$\moth \setbox\zo\hbox{\tick}%
  \hskip 3pt\hbox to 0pt{$\tick$\hss}\hrulefill \hbox to 7.5pt{$\tick$\hss}$}

\def\dtick{\leaders\hrule height .34pt depth 0.5ex \hskip 0.5pt}
\def\downboxfill{$\moth \setbox\zo\hbox{\dtick}%
  \hskip 2pt\hbox to 0pt{$\dtick$\hss}\hrulefill \hbox to 2pt{$\dtick$\hss}$}

\def\bec{\begin{center}}
\def\ec{\end{center}}

\def\cF{{\cal F}}

\def\cH{{\cal H}}

\def\be{\begin{equation}}
\def\ee{\end{equation}}
\def\bea{\begin{eqnarray}}
\def\eea{\end{eqnarray}}
\def\ba{\begin{array}}
\def\ea{\end{array}}

\usepackage{color}
\definecolor{red}{rgb}{1,0,0}
\definecolor{lred}{rgb}{0.3,0,0}
\definecolor{green}{rgb}{0,0.6,0}
\definecolor{blue}{rgb}{0,0,1}
\definecolor{violet}{rgb}{0.8,0,0.8}

\begin{document}

\begin{titlepage}
\rightline{}
\rightline{December 2016}
\begin{center}
\vskip .6cm
{\LARGE \bf { Generalized IIB supergravity from\\[1ex]
exceptional field theory}}\\
\vskip 1cm
{\large {Arnaud Baguet, Marc Magro and Henning Samtleben}}

{\it 
Univ Lyon, Ens de Lyon, Univ Claude Bernard, CNRS,\\
Laboratoire de Physique, F-69342 Lyon, France} \\[1ex]
{\tt arnaud.baguet, marc.magro, henning.samtleben@ens-lyon.fr}

\vskip 1cm
{\bf Abstract}
\end{center}

\vskip 0.2cm

\noindent
\begin{narrower}

The background underlying the 
$\eta$-deformed $AdS_5\times S^5$ sigma-model is known to 
satisfy a generalization
of the IIB supergravity equations. 
Their solutions are related by T-duality to solutions of type IIA supergravity 
with non-isometric linear dilaton.
We show how the generalized IIB supergravity equations
can be naturally obtained from exceptional field theory.
Within this manifestly duality covariant formulation of maximal supergravity,
the generalized IIB supergravity equations emerge upon
imposing on the fields a simple Scherk-Schwarz ansatz
which respects the section constraint.

\end{narrower}

\vskip 1.5cm

\end{titlepage}



\section{Introduction}

Integrability plays a key role in the study of AdS/CFT correspondence 
\cite{Maldacena:1997re,Gubser:1998bc,Witten:1998qj} between 
Type IIB superstring theory on the $AdS_5 \times S^5$ background
 and
the maximally supersymmetric Yang-Mills gauge theory in four
dimensions  (see \cite{Beisert:2010jr} for a review). The former superstring theory 
is described by the Metsaev-Tseytlin action \cite{Metsaev:1998it}. 
This action is the sum of two terms: a kinetic one and an exact Wess-Zumino-like one. The  
value of the coefficient in front of the  Wess-Zumino term is fixed by requiring invariance 
under $\kappa$-symmetry. There is a very nice interplay between   $\kappa$-symmetry 
and integrability. Indeed,  the  value of the aforementioned coefficient which guaranties 
$\kappa$-symmetry also ensures integrability. This means that the equations of motion   
admit  a zero-curvature formulation in terms of a Lax pair \cite{Bena:2003wd} 
(see \cite{Arutyunov:2009ga} for a review). 

\medskip

Recall that $\kappa$-symmetry is a fermionic gauge invariance.  It has been shown in
\cite{Delduc:2013qra,Kawaguchi:2014qwa,Hollowood:2014qma} that it is possible 
to deform the  
$AdS_5 \times S^5$ superstring while maintaining both properties of integrability and invariance 
under a local fermionic symmetry. The $\sigma$-model constructed in 
\cite{Delduc:2013qra} has been dubbed $\eta$-deformation. The   one  
constructed in \cite{Kawaguchi:2014qwa}  is sometimes also designated as
$\eta$-deformation in the literature. These theories are also called 
inhomogeneous/homogeneous Yang-Baxter deformations because their definitions involve a $R$-matrix, 
which is a solution of an inhomogeneous/homogeneous equation respectively. The latter 
equations are the modified classical Yang-Baxter equation and   the 
classical Yang-Baxter equation respectively.  The deformation proposed in \cite{Hollowood:2014qma} 
  has been called  $\lambda$-deformation of the $AdS_5 \times S^5$ superstring.
  The actions  which have been put forward  in \cite{Delduc:2013qra,Kawaguchi:2014qwa} 
 are deformations of the Metsaev-Tseytlin action while the action associated with the 
$\lambda$-deformation \cite{Hollowood:2014qma} 
 appears  as a deformation of the non-abelian T-dual of the 
$AdS_5 \times S^5$ superstring $\sigma$-model. Inhomogeneous 
Yang-Baxter deformations are related \cite{Vicedo:2015pna,Hoare:2015gda,Sfetsos:2015nya,
Klimcik:2015gba} by Poisson-Lie T-duality \cite{Klimcik:1995ux,Klimcik:1995dy} 
and analytic continuation to $\lambda$-deformations. 
 Such deformations generalise techniques 
applied to  
the principal chiral model and  symmetric space $\sigma$-models 
\cite{Klimcik:2002zj,Klimcik:2008eq,Delduc:2013fga,Sfetsos:2013wia,Hollowood:2014rla}. 
  These deformations, 
  some further generalisation as well as lower dimensional cases 
have been thoroughly explored in many articles 
(see references for instance in the thesis \cite{Borsato:2016hud} and the reviews \cite{
Matsumoto:2014cja,Thompson:2015lzd}). 

\medskip

It has been shown recently that the  actions of the $\eta$- 
  and $\lambda$-deformations 
can be put on the Green-Schwarz form and that 
their fermionic local symmetry corresponds to   standard $\kappa$-symmetry \cite{Borsato:2016ose}. 
On a more general and fundamental level, a full understanding of these models 
from a supergravity perspective 
 has required   to revisit and underpin precisely  the relationship 
 between $\kappa$-symmetry and supergravity. This has been achieved a short while ago 
\cite{Wulff:2016tju} and may be summarised as follows. Conditions to have classical 
$\kappa$-symmetry of Green-Schwarz $\sigma$-model are not equivalent\footnote{See also 
\cite{Mikhailov:2012id} for a related discussion in the pure spinor formulation.} to Type IIB supergravity 
equations but to the generalisation of these equations which has been 
derived in \cite{Arutyunov:2015mqj,Wulff:2016tju}. 
 An 
equivalent statement is that classical $\kappa$-symmetry does not imply two-dimensional 
Weyl invariance for the $\sigma$-model but only scale invariance. 

The 
 generalized supergravity equations 
  comprise a Killing vector field 
$K$ and ordinary Type IIB supergravity equations are recovered whenever $K$ vanishes. 
For instance, the background  associated with the $\lambda$-deformation  
has no isometry. This explains why the $\lambda$-deformation  defines generically  
\cite{Borsato:2016ose}  
a Type IIB supergravity background (see also \cite{Borsato:2016zcf,Chervonyi:2016ajp,Chervonyi:2016bfl}  
and \cite{Sfetsos:2014cea,Demulder:2015lva}). On the 
contrary, the
Arutyunov-Borsato-Frolov (ABF) background 
\cite{Arutyunov:2013ega,Arutyunov:2015qva,Borsato:2016hud}, which 
 corresponds to the $\eta$-deformation in \cite{Delduc:2013qra}, is a solution 
 of the generalized supergravity equations \cite{Arutyunov:2015mqj}. A general 
 discussion,  including 
 the case of homogeneous Yang-Baxter deformations, and   examples of the latter 
  may be found in \cite{Borsato:2016ose} and in 
  \cite{Kyono:2016jqy,Hoare:2016ibq,Hoare:2016hwh,Orlando:2016qqu} respectively.

The presence of a Killing vector field plays an important role. Indeed, solutions 
of generalized Type II supergravity are related by formal T-duality to ordinary supergravity solutions. 
This may be illustrated by two examples \cite{Arutyunov:2015mqj}. 
Consider firstly the  Hoare-Tseytlin (HT) Type IIB background   described in 
\cite{Hoare:2015wia}. It consists of a metric $\widetilde{G}$, a (imaginary)  5-form 
$\widetilde{F}_5$ and a dilaton 
$\widetilde{\phi}$. The  metric and the product 
$\widetilde{{\cal F}}_5 = e^{\widetilde{\phi}} \widetilde{F}_5$ are invariant under a $U(1)^6$ 
isometry. However, the dilaton contains 
a linear non-isometric term, which breaks four of these isometries.  Applying   a formal T-duality 
  in all   isometric directions on $\widetilde{G}$ and $\widetilde{{\cal F}}_5$ leads 
  \cite{Hoare:2015wia} to the ABF background. The second example is simpler and explains  the mechanism at work. We 
reproduce it from   \cite{Arutyunov:2015mqj}.  
Consider the equations  
 \begin{equation} \label{ord-simp}
 \widetilde{R}_{mn} + 2 {\widetilde \nabla}_m {\widetilde \nabla}_n 
 {\widetilde \phi}=0 \;,\qquad \mbox{and} \qquad 
   {\widetilde R}  + 4\, {\widetilde \nabla}^2 {\widetilde \phi}   - 4  \partial_m {\widetilde \phi}
    \partial^m {\widetilde \phi}=0\;,
  \end{equation}
for a metric  and dilaton   of the form 
\begin{equation*}
{ \widetilde d s}^2 = e^{2a(x)} [d \widetilde{y} + A_\mu(x) dx^\mu]^2 + g_{\mu\nu} (x) dx^\mu dx^\nu   \qquad \mbox{and} \qquad 
 \widetilde{\phi}(\widetilde{y},x) = - c \widetilde{y} + f(x),
\end{equation*}
with $c$ a constant. The metric has an isometry which is broken by the linear term in the dilaton.
Applying T-duality to the metric leads to the metric and B-field:
 \begin{equation} \label{met-b-ex-dual}
ds^2= e^{-2a(x)} dy^2 + g_{\mu\nu} (x) dx^\mu dx^\nu \qquad \mbox{and} \qquad 
 B=  A_\mu(x) \, dy \wedge dx^\mu .
\end{equation}
Introduce then 
the vector $X = K + Z$, where the non-zero components 
of the Killing vector $K$ and $Z$ are  
$K_{y}= c \, e^{- 2 a(x)}$ and $Z_\mu = \partial_\mu(f-a)  - 
B_{\mu y} K^{y}$.  Then, equations \eqref{ord-simp} become for $X$ and 
the metric and the field strength $H$ of the B-field in  \eqref{met-b-ex-dual} 
\begin{equation} \label{gen-simp}
 {R}_{mn} +  {\nabla}_m X_n + {\nabla}_n X_m =0 ,\qquad 
 \mbox{and} \qquad    R - \frac{1}{12} \,H^2_{mnp} + 4\, \nabla_m X^m - 4 X_m X^m =0.
\end{equation}
When $c$ vanishes,   $K$ vanishes as well. Then $X_m =Z_m= 
\partial_m (f-a) \equiv 
\partial_m \phi$ and therefore 
  equations \eqref{gen-simp}   generalise  equations \eqref{ord-simp}. 
 
The generalized IIB equations from~\cite{Arutyunov:2015mqj} may be obtained by applying 
an analogous generalized T-duality to the full field equations of IIA supergravity again 
with non-isometric linear dilaton. This suggests that they should have
a natural place in the manifestly duality covariant formulation of maximal supergravities -- the
so-called exceptional field theories (ExFT)~\cite{Hohm:2013pua,Hohm:2013vpa,Hohm:2013uia}.
These give a U-duality covariant unified description of IIA and IIB supergravity on a formally enhanced
exceptional space-time. According to the solution of the section constraint which selects the physical
coordinates, their field equations reproduce the standard IIA and the IIB field equations, respectively.
We will show in this article, that also the generalized IIB supergravity equations are straightforwardly
obtained from exceptional field theory upon choosing a particular solution of the section constraint.

Our framework for this article is the E$_{6(6)}$ covariant exceptional field theory 
from~\cite{Hohm:2013pua,Hohm:2013vpa}, most adapted to the $5+5$ split of the IIB coordinates for 
the undeformed $AdS_5\times S^5$ background. In E$_{6(6)}$ ExFT, the $AdS_5$ coordinates $\{x^\mu\}$
remain external coordinates while the 5 coordinates $\{y^a\}$ describing the
$S^5$-geometry of the undeformed background, are internal and
formally embedded into 27 coordinates $\{Y^M\}$ transforming in
the fundamental representation of the group E$_{6(6)}$.
Dependence of all fields on the 27 coordinates is heavily constrained by the section constraint stating that
\bea
d^{KMN} \partial_M \otimes \partial_N ~\equiv~0
\;,
\label{section0}
\eea
whenever the derivatives act on the various fields of the theory. Here, $d^{KMN}$ denotes the cubic invariant 
tensor of E$_{6(6)}$.
The field content of this ExFT is given in terms of E$_{6(6)}$ covariant fields, most notably a group-valued 
symmetric $27\times 27$ matrix ${\cal M}_{MN}$ parametrizing the coset space ${\rm E}_{6(6)}/{\rm USp}(8)$\,.
The type IIB theory is recovered upon breaking E$_{6(6)}$ down to its subgroup
${\rm SL}(5)\times{\rm SL}(2)\times{\rm GL}(1)_{\rm IIB}$, such that
\bea
{\bf 27} &\longrightarrow&
(5,1)_{+4} + (5',2)_{+1} +(10,1)_{-2} + (1,2)_{-5}
\;,\nonumber\\
\left\{Y^M\right\} &\longrightarrow& \{y^a\,,\; \tilde{y}_{a\alpha}\,,\;  \tilde{y}^{ab}\,,\;  \tilde{y}_{\alpha} \}
\;,\qquad
a=1, \dots, 5\;,\quad \alpha=\pm
\;.
\label{breakIIB}
\eea
Restricting the dependence of all fields to the 5 coordinates $\{y^a\}$ 
of highest grading under the ${\rm GL}(1)_{\rm IIB}$ provides a solution to (\ref{section0}).
Decomposing the matrix ${\cal M}_{MN}$ into its various blocks according to (\ref{breakIIB}) allows
to identify the explicit dictionary to the various components of the IB fields~\cite{Baguet:2015xha}.
The field equations of ExFT in this case reproduce the equations of standard IIB supergravity.

We will show in this article that generalized IIB supergravity equations are obtained upon choosing a different
solution of the section condition. Recall that among the ingredients of the generalized equations
features a Killing vector field $K^a$ which after splitting $\{y^a\}=\{y^i, y^*\}$, $(i=1, \dots, 4)$, 
we choose along the fixed direction $K^a=\delta_*^a$,
such that $\partial_*\Phi=0$ for all fields of the theory.
In this case, the section constraint (\ref{section0}) admits a different (and inequivalent) solution in 
which we restrict the dependence of all fields to the subset $\{y^i, \tilde{y}_{*+}\}$,
i.e.\ allow for an additional dependence on the coordinate $\tilde{y}_{*+}$.
We will show that upon imposing a particular dependence on $\tilde{y}_{*+}$ (in terms of a generalized
Scherk-Schwarz reduction ansatz), the field equations of exceptional field theory precisely reproduce the 
generalized IIB supergravity equations. 

We should stress that the existence of an inequivalent solution to the section constraint does not come
as a surprise but this choice of coordinates 
is equivalent (after rotation) to selecting the IIA coordinates $\tilde{y}_{a+}$ in (\ref{breakIIB}),
in which case ExFT reproduces the standard IIA theory. This precisely amounts to the fact that generalized IIB supergravity
can be obtained via T-duality from a sector of IIA supergravity. Since the framework of exceptional field theory
is manifestly duality covariant, we do not have to go through this duality explicitly but can simply absorb its effect
into a rotation of the extended coordinates. Evaluating ExFT according to the IIB dictionary however in the new
coordinates $\{y^i, \tilde{y}_{*+}\}$ with proper Scherk-Schwarz ansatz in $\tilde{y}_{*+}$ 
then directly yields the generalized IIB supergravity equations.

The rest of the article is organized as follows: in section 2 we collect the generalized IIB supergravity equations.
In section 3 we start with a brief review of the relevant E$_{6(6)}$ exceptional field theory and present the
generalized Scherk-Schwarz ansatz that governs the $\tilde{y}_{*+}$-dependence of the fields. 
We work out the ExFT field equations with this ansatz and show that they yield the generalized IIB supergravity equations.

\bigskip

While we were writing up these results, Sakatani, Uehara, and Yoshida submitted the interesting paper \cite{Sakatani:2016fvh} to the archive which in a similar spirit relates the NS-NS truncation of the generalized type IIB supergravity equations to the $O(d, d)$ covariant double field theory~\cite{Siegel:1993th,Hull:2009mi,Hohm:2010jy,Hohm:2010pp}.

\section{Generalized IIB supergravity}
\label{sec:genIIB}

\subsection{Generalized field equations and Bianchi identities}

We recall in this section the bosonic generalized IIB supergravity equations which have been derived 
in \cite{Arutyunov:2015mqj}.  Their fermionic completion has been found in~\cite{Wulff:2016tju}. 
The equations are expressed in string frame. The equations 
for the metric $G_{mn}$ and the $B$-field $B_{mn\,+}$ are
\begin{subequations} \label{set 1}
\begin{align}
R_{mn} -  \frac{1}{4} H_{mpq} H_{n}{}^{pq} - {\cal T}_{mn} + \nabla_m X_n + \nabla_n X_m &=0, \label{eom metric}\\
\frac{1}{2} \,\nabla^p H_{pmn} + \frac{1}{2}  {\cal F}^{p} {\cal F}_{pmn} + 
\frac{1}{12}\, {\cal F}_{mnpqr} {\cal F}^{pqr} - X^p H_{pmn} - \partial_m X_n + \partial_n X_m  &=0, 
\label{eom H}\\
  R - \frac{1}{12} \,H^2_{mnp} + 4\, \nabla_m X^m - 4 X_m X^m =0\;,
   \label{eom dil}
\end{align}
where $\nabla_m$ denotes the space-time covariant derivative, $R_{mn}$ the Ricci tensor, $R$ the Ricci scalar and 
 $H_{mnp}=3\,\partial_{[m}B_{np]\,+}$   the field strength of the NS-NS $B$-field.  
  The R-R fields enter via the currents ${\cal F}_{m_1 \cdots m_n} $ 
  and contribute to the stress tensor in (\ref{eom metric}) via
\begin{equation}
{\cal T}_{mn} = \frac{1}{2} \,{\cal F}_m {\cal F}_n + \frac{1}{4}\, {\cal F}_{mpq} {\cal F}_{n}{}^{pq} +
\frac{1}{4\times 4!}\, {\cal F}_{mpqrs} {\cal F}_{n}{}^{pqrs} -  \frac{1}{4} \,G_{mn} \Bigl(
 {\cal F}_p {\cal F}^p + \frac{1}{6}\, {\cal F}_{pqr} {\cal F}^{pqr} \Bigr)\;.
\end{equation}
\end{subequations}
The equations (\ref{eom metric})--(\ref{eom dil}) are based on the existence of a Killing vector field
$K$ and an additional vector field $Z$ with $K^m Z_m=0$, which enter the field equations
in the combination $X\equiv K+Z$.
 The vector field $Z$ satisfies the Bianchi type equations
 \bea
  \partial_m Z_n -\partial_n Z_m + K^p H_{pmn} &=&0\;. \label{iso 3form H}
  \eea
The ordinary type IIB equations are recovered in the limit where $K =0$ such
that $Z$ can be integrated to the dilaton field $Z_m = \partial_m \phi$\,. 

In the R-R sector, the generalized dynamical equations  for the 
field strengths ${\cal F}_{m_1 \cdots m_n} $ are given by
 \begin{subequations} \label{ef1-ef5}
 \begin{align}
& \nabla^m {\cal F}_m - Z^m {\cal F}_m -   \frac{1}{6}\, H^{mnp} {\cal F}_{mnp} =0, \qquad K^m {\cal F}_m = 0,
 \label{ef1} \\
& \nabla^p {\cal F}_{pmn} - Z^p {\cal F}_{pmn} -   \frac{1}{6} \,H^{pqr}{\cal F}_{mnpqr} - 
 (K \wedge {\cal F}_1)_{mn}= 0, \label{ef3}\\
& \nabla ^r {\cal F}_{rmnpq} - Z^r {\cal F}_{rmnpq} +   \frac{1}{36}\,
 \varepsilon_{mnpqrstuvw} H^{rst} {\cal F}^{uvw} - 
 (K \wedge {\cal F}_3)_{mnpq} =0,  \label{ef5} 
 \end{align}
\end{subequations}
while their modified Bianchi identities can be cast into the compact form
\bea
 d {\cal F}_{2n+1}- Z \wedge {\cal F}_{2n+1} + H_3 \wedge  {\cal F}_{2n-1}&=&
 \star\left(K\wedge \star{\cal F}_{2n+3}\right)
 \;.
 \label{BianchiF}
\eea
The Bianchi identities extend to the dual field strengths $-{\cal F}_7\equiv\star{\cal F}_3$ 
and ${\cal F}_9\equiv\star{\cal F}_1$. Furthermore, the selfduality property
${\cal F}_{rmnpq}=\star{\cal F}_{rmnpq}$ of the five form continues to hold in the modified theory,
relating its Bianchi identity and field equation.
In the following, for simplicity of the formulas, we will often choose coordinates such that the Killing vector field
points in a given direction $K^m =\delta^m_*$.

\subsection{Solution of the Bianchi identities}

It has been noted in \cite{Arutyunov:2015mqj} that equation (\ref{iso 3form H}) for the new vector $Z_m$ may be
interpreted as  a modified ``dilaton Bianchi identity'' and locally integrated into
\bea
Z_m &=&  \partial_m \phi + K^p B_{pm\, +} ~=~ \partial_m \phi - B_{m*\,+}
\;.
\label{ZDP}
\eea
We will in the following stay in this picture and understand the combination $\partial_m \phi - B_{m*\,+}$
as a derivative $D_m \phi$ on the dilaton that is covariantized in a suitable sense.
As a related observation, one may straightforwardly check that the modified Bianchi
identities (\ref{BianchiF}) satisfied by the R-R field strengths 
allow for an explicit integration into ${\cal F}=e^\phi\,F$ with
\bea
{F}_m &=&
\partial_m \chi + B_{m*\,+}\,\chi  +B_{m*\,-}  ~\equiv~
D_m \chi 
\;, \nonumber\\
{F}_{pmn} &=& 
3\,\partial_{[p}B_{mn]\,-} +
\frac32\,B_{[p|*+|} B_{mn]\,-} - \frac32\,B_{[p|*-|} B_{mn]\,+} +C_{pmn*}+ \chi\,H_{pmn}
\;,\nonumber\\
{F}_{mnpqr} &=& 
5\,\partial_{[m}C_{npqr]}+5\,B_{[m|*+|} C_{npqr]}
-15\,  B_{[mn\,|+|} \partial_{p}B_{qr]\,-}\nonumber\\
&& {}\qquad
-15\, B_{[mn\,|+|}B_{p|*\,+|}B_{qr]\,-}
+15\,  B_{[mn\,|-|}\partial_{p}B_{qr]\,+}
+ C_{mnpqr*\,+}
\;,
\nonumber\\
F_{mnpqrst}&=&
7\,\partial_{[m}C_{npqrst]\,+}+7\,B_{[m*+}C_{npqrst]\,+}+35\,C_{[mnpq}H_{rst]}
\nonumber\\
&&{}
\qquad -105\,B_{[mn|\,+|}B_{pq|\,-|}H_{rst]}+C_{mnpqrst*}\;.
\label{FFF}
\eea
All the terms, carrying indices $`{}_*{}`$ represent the deformations from the standard IIB expressions.
Again, in the following we will assign them a natural interpretation as the connection terms of covariantized derivatives,
non-abelian field strengths and the St\"uckelberg type couplings among $p$-forms. 
These additional couplings precisely match the structure of general nine-dimensional gauged 
supergravities\cite{Bergshoeff:2002nv,FernandezMelgarejo:2011wx} (recall that due to the existence
of a Killing vector field, we are effectively describing a nine-dimensional theory).
More precisely, equations (\ref{FFF}) can be viewed as resulting from a gauging of nine-dimensional maximal supergravity
in which a linear combination of the Cartan subgroup of the ${\rm SL}(2)_{\rm IIB}$ and the trombone symmetry
which scales every field according to its Weyl weight has been gauged. The component $B_{m*\,+}$ 
of the ten-dimensional NS-NS two-form serves as a gauge field.\footnote{
To be precise, also a nilpotent generator of ${\rm SL}(2)_{\rm IIB}$ is gauged with the component $B_{m*\,-}$ 
serving as the associated gauge field.}
An important consequence
is the following. According to (\ref{ZDP}), the dilaton $\phi$ is charged under the new local gauge symmetry.
Translation to the Einstein frame via
\bea
G_{mn}^{\rm st} &=& e^{\phi/2}\,G_{mn}^{\rm E} 
\;,
\eea
thus implies that the metric in the Einstein frame is also charged. Translation of the Einstein field equations (\ref{eom metric}) 
into the Einstein frame thus induces field equations which feature a covariantized Ricci tensor in the sense that all
derivatives in its definition are replaced by properly covariantized ones. In particular, the Riemann tensor is calculated
as curvature of the connection
\bea
\hat\Gamma_{mn}{}^p &\equiv& \frac12\,G^{pq}\left(
D_{m} G_{nq}+D_{n} G_{mq} -D_q G_{mn}
\right)\;,\quad 
D_p G_{mn}\equiv \partial_p G_{mn} +\frac12\,B_{p*\,+}\,G_{mn}
\;.
\label{metricTromb}
\eea
This is the generic structure of supergravities in which the trombone symmetry is gauged~\cite{LeDiffon:2008sh}.
Upon transition to the Einstein frame, we may also regroup the field equations for NS-NS and R-R two-form
(\ref{eom H}) and (\ref{ef3})
into the manifestly ${\rm SL}(2)$ covariant form
\bea
D_{p}\left(F^{pmn\,\alpha}\,m_{\alpha\beta}\right)-\frac16\,F^{mnpqr}\,F_{pqr}{}^\alpha \,\varepsilon_{\alpha\beta}
&=& J^{mn}{}_{\beta}\;,
\label{DFJ}
\eea
with the ${\rm SL}(2)$ doublet 
$F_{mnp\,\pm}=\{H_{mnp},F_{mnp}-\chi\,H_{mnp}\}$, 
and the dilaton/axion matrix $m_{\alpha\beta}$ parametrized as
\bea
m_{\alpha\beta}&=&
\begin{pmatrix}
e^{\phi} & -e^\phi\chi \\
-e^\phi\chi & e^\phi\chi^2+e^{-\phi}
\end{pmatrix}
\;.
\eea
The current on the r.h.s.\ of (\ref{DFJ}) is given by the ${\rm SL}(2)$ doublet
\bea
J^{mn}{}_{\pm}&=&\left\{2\,e^{2\phi}\,K^{[m}F^{n]}
\,,\;-4\,e^{2\phi}\,\nabla^{[m}K^{n]}-2\,\chi \,e^{2\phi}\, K^{[m}F^{n]} \right\}
\;,
\eea
in terms of the Killing vector field $K^m$ and the current $F_m=D_m\chi$\,.
We will in the following recover the non-abelian field-strengths (\ref{FFF}) 
from a particular Scherk-Schwarz ansatz in exceptional field theory.

\section{Exceptional field theory}

\subsection{E$_{6(6)}$ exceptional field theory}

In this  section, we give a brief review of the exceptional field theory associated with the group E$_{6(6)}$ 
into which we will in the following embed the generalized IIB supergravity equations.
We refer to~\cite{Hohm:2013vpa,Musaev:2014lna,Baguet:2015xha} for details of the theory.
The bosonic field content of E$_{6(6)}$ exceptional field theory reflects the
field content of maximal $D=5$ supergravity, and is given by 
\bea
\left\{g_{\mu\nu}\,,\; {\cal M}_{MN},\; {\cal A}_\mu{}^M\,,\; {\cal B}_{\mu\nu\,M} \right\}
\;,
\label{fieldcontent}
\eea
with indices $\mu, \nu = 0, \dots, 4,$ and $M = 1, \dots, 27$. 
The symmetric matrices $g_{\mu\nu}$
and ${\cal M}_{MN}$ define the external and internal metrics, respectively,
the latter parametrizing the coset space ${\rm E}_{6(6)}/{\rm USp}(8)$\,.
In addition to their dependence on the five external coordinates $\{x^\mu\}$, 
the fields~(\ref{fieldcontent}) formally depend on
27 internal coordinates $\{Y^M\}$ transforming in the fundamental representation of ${\rm E}_{6(6)}$.
The latter dependence is strongly restricted by the ${\rm E}_{6(6)}$ covariant section 
condition~\cite{Berman:2010is,Coimbra:2011ky}
(on any two fields or gauge parameters $A, B$)
\bea
  d^{KMN}\,\partial_M \partial_N A &=& 0\;, \qquad  d^{KMN}\,\partial_M A\, \partial_N B ~=~ 0 \,,
   \label{sectioncondition}
 \eea  
 with $d^{KMN}$ denoting the totally symmetric cubic invariant of ${\rm E}_{6(6)}$.
The constraints (\ref{sectioncondition}) admit two inequivalent solutions, in which the fields
depend on a subset of six or five of the internal coordinates.
Upon implementing one of these solutions, the field equations of exceptional field theory
reproduce full $D=11$ supergravity and ten-dimensional IIB supergravity, respectively.

The field equations of exceptional field theory are most compactly obtained from variation of a Lagrangian
 \bea
 \sqrt{|g|}{}^{-1}\, {\cal L}_{\rm ExFT} & = & \widehat{R}
 +\frac1{24}\,g^{\mu\nu}\,{D}_{\mu}{\cal M}_{MN}\,{D}_{\nu}{\cal M}^{MN}
 -\frac14\,{\cal F}_{\mu\nu}{}^M{\cal F}^{\mu\nu\,N}\,{\cal M}_{MN}
\nonumber\\
&&{}
+\sqrt{|g|}{}^{-1}\, {\cal L}_{\rm top} - V_{\rm pot}
  \;. 
   \label{EFTaction}
 \eea 
The first term formally takes the same form as the five-dimensional Einstein-Hilbert term, 
 where in the definition of the Ricci scalar $\widehat{R}$ all external derivatives are covariantized w.r.t.\
 the action of internal diffeomorphisms (under which the external metric transforms as a weighted scalar)
 \bea
 \partial_\mu g_{\nu\rho}~\longrightarrow~
 D_\mu g_{\nu\rho} &\equiv& \partial_\mu g_{\nu\rho} - {\cal A}_\mu{}^M \partial_M g_{\nu\rho}
  -\frac23\,\partial_M {\cal A}_\mu{}^M\,g_{\nu\rho}
  \;.
  \label{Dgg}
 \eea
The second term in (\ref{EFTaction}) is a gauged coset space sigma-model ${\rm E}_{6(6)}/{\rm USp}(8)$
with derivatives $D_\mu\equiv \partial_\mu - {\cal L}_{{\cal A}_\mu}$ 
covariantized under the action of generalized internal 
diffeomorphisms~\cite{Coimbra:2011ky}
\bea\label{genLie}
{\cal L}_\Lambda {\cal M}_{MN} &=&  \Lambda^K \partial_K {\cal M}_{MN} +2\,\partial_{(M} \Lambda^K {\cal M}_{N)K}
  -10\, \partial_K\Lambda^L\,d^{KPR} d_{RL(M}  {\cal M}_{N)P}
  \nonumber\\
  &&{}
  +\frac23\, \partial_P\Lambda^P\,  {\cal M}_{MN}
  \;.
\eea
The third term in (\ref{EFTaction}) describes a Yang-Mills type kinetic term for the 
27 gauge vectors ${\cal A}_\mu{}^M$ whose non-abelian structure again is induced
by the structure of internal diffeomorphisms 
\bea
{\cal F}_{\mu\nu}{}^M &=&  2\, \partial_{[\mu} {\cal A}_{\nu]}{}^M  -2\,{\cal A}_{[\mu}{}^K \partial_K {\cal A}_{\nu]}{}^M 
+10\, d^{MKR}d_{NLR}\,{\cal A}_{[\mu}{}^N\,\partial_K {\cal A}_{\nu]}{}^L
\nonumber\\
&&{}+ 10 \, d^{MNK}\,\partial_K {\cal B}_{\mu\nu\,N}
\;.
\label{FYM}
\eea
The St\"uckelberg-type coupling to the 27 two-forms ${\cal B}_{\mu\nu\,N}$ induces a modification
of the Bianchi identities as
\bea
3\,D_{[\mu} {\cal F}_{\nu\rho]}{}^M &=& 
10 \, d^{MNK}\,\partial_K {\cal H}_{\mu\nu\rho\,N}
\;,
\label{defH}
\eea
with the 3-form field strength ${\cal H}_{\mu\nu\rho\,N}$ defined by this equation
(up to terms that vanish under projection with $d^{MNK}\,\partial_K$).
The term ${\cal L}_{\rm top}$ in (\ref{EFTaction}) refers to a Chern-Simons-type topological term
which is such that the field equations obtained by varying the full Lagrangian w.r.t.\ the 
two-forms ${\cal B}_{\mu\nu\,M}$
give rise to the first order duality equations
\bea\label{dualityrel}
d^{MNK}\partial_K  \left( \sqrt{|g|}\,{\cal M}_{NL}\, {\cal F}^{\mu\nu L}
 +\frac16\,\sqrt{10}\,  \varepsilon^{\mu\nu\rho\sigma\tau}\,
  {\cal H}_{\rho\sigma\tau N}\right) &=& 0\;,
\eea 
relating the field strengths of vector fields and two-forms.
Finally, the last term in (\ref{EFTaction}) is the 
`scalar potential' that involves only internal derivatives $\partial_M$ 
whose explicit form can be found in \cite{Hohm:2013vpa}.
Its form is uniquely determined by invariance under internal generalized diffeomorphisms.

\subsection{Section constraints and IIA/IIB/generalized supergravity}

The section condition (\ref{sectioncondition}) is solved by restricting the internal coordinate 
dependence of all fields to properly chosen subsets of coordinates.
Breaking E$_{6(6)}$ down to its subgroup
${\rm SL}(5)\times{\rm SL}(2)\times{\rm GL}(1)_{\rm IIB}$
according to
\bea
{\bf 27} &\longrightarrow&
(5,1)_{+4} + (5',2)_{+1} +(10,1)_{-2} + (1,2)_{-5}
\;,\nonumber\\
\left\{Y^M\right\} &\longrightarrow& \{y^a\,,\; \tilde{y}_{a\alpha}\,,\;  \tilde{y}^{ab}\,,\;  \tilde{y}_{\alpha} \}
\;,\qquad
a=1, \dots, 5\;,\quad \alpha=\pm
\;,
\label{breakIIB1}
\eea
and restricting all fields to depend only on the 5 coordinates $\{y^a\}$ of highest grading under ${\rm GL}(1)_{\rm IIB}$
solves the conditions (\ref{sectioncondition}). The ExFT field equations derived from (\ref{EFTaction}) then reproduce the 
IIB theory after decomposing the ExFT fields (\ref{fieldcontent}) according to (\ref{breakIIB1}) and properly
translating the various blocks into the various components of the IIB fields \cite{Baguet:2015xha}.
In particular, the scalar matrix ${\cal M}_{MN}$ decomposes into
\bea
{\cal M}_{KM} &=& \left(
\begin{array}{cccc}
{\cal M}_{ac}&{\cal M}_a{}^{c\beta}&{\cal M}_{a,cd}&{\cal M}_{a}{}^{\beta}\\
{\cal M}^{a\alpha}{}_{c}&{\cal M}^{a\alpha,}{}^{c\beta}&{\cal M}^{a\alpha}{}_{cd}&{\cal M}^{a\alpha,}{}^{\beta}\\
{\cal M}_{ab,c}&{\cal M}_{ab}{}^{c\beta}&{\cal M}_{ab,cd}&{\cal M}_{ab}{}^{\beta}\\
{\cal M}^\alpha{}_{c}&{\cal M}^{\alpha,c\beta}&{\cal M}^\alpha{}_{cd}&{\cal M}^\alpha{}^{\beta}
\end{array}
\right)
\;,
\label{M2B}
\eea
where the explicit form of the blocks is obtained by evaluating the matrix 
${\cal M}_{MN}={\cal V}_M{}^A{\cal V}_N{}^A$ from a vielbein ${\cal V}_M{}^A$ given by the 
product of matrix exponentials
\bea
{\cal V}_{\rm IIB} &\equiv& 
{\rm exp}\left[\varepsilon^{abcde}\,c_{abcd} \, t_{(+4)\,e}\right] \,
{\rm exp}\left[b_{ab}{}^\alpha\,t_{(+2)}{}_{\alpha}^{ab}\right]\,
{\cal V}_5\,{\cal V}_2\,
{\rm exp} \left[\Phi\, t_{\rm IIB}\right]
\;,
\label{V27AB}
\eea
with the relevant $\mathfrak{e}_{6(6)}$ generators $t_{\rm IIB}$, $t_{(+2)}{}_{\alpha}^{ab}$, $t_{(+4)\,a}$
and their coefficients originating from the IIB metric, two-form and four-form, respectively.
The matrices ${\cal V}_5$, ${\cal V}_2$ represent the ${\rm SL}(5)\times{\rm SL}(2)$ factor
of the vielbein, related to the internal metric and the IIB dilaton/axion matrix, respectively.
Similarly, vectors and two-forms are decomposed as (\ref{breakIIB1})
\bea
\left\{{\cal A}_\mu{}^M\right\} &\longrightarrow& \{{\cal A}_\mu{}^a\,,\; {\cal A}_\mu{\,}_{a\alpha}\,,\;  {\cal A}_\mu{}^{ab}\,,\;  {\cal A}_\mu{\,}_{\alpha} \}\;,\nonumber\\
\left\{{\cal B}_{\mu\nu}{\,}_M\right\} &\longrightarrow& \{{\cal B}_{\mu\nu}{\,}_a\,,\; {\cal B}_{\mu\nu}{}^{a\alpha}\,,\;  {\cal B}_{\mu\nu}{\,}_{ab}\,,\;  {\cal B}_{\mu\nu\,\alpha} \}\;.
\eea

In contrast, the IIA theory is recovered, if the physical coordinates are identified with the $\{ \tilde{y}_{a+}\}$
in the decomposition (\ref{breakIIB1}) (which explicitly breaks the ${\rm SL}(2)$ factor), the ExFT fields
are decomposed accordingly and translated into the IIA fields. E.g.\ in this case, the proper parametrization of
the matrix ${\cal M}_{MN}={\cal V}_M{}^A{\cal V}_N{}^A$ is obtained via a vielbein ${\cal V}_M{}^A$ 
\bea
{\cal V}_{\rm IIA} &\equiv& 
{\rm exp}\left[\varphi \, t_{(+5)}\right] \,
{\rm exp}\left[\,c_{abc} \, t_{(+3)}^{abc}\right] \,
{\rm exp}\left[b_{ab}{}\,t_{(+2)}^{ab}\right]\,
{\rm exp}\left[c_{a}{}\,t_{(+1)}^{a}\right]\,
{\cal V}_5\,{\rm exp} \left[\phi\, t_{0}+\Phi\, t_{\rm IIA}\right]
\;,
\nonumber\\
\label{VIIA}
\eea
with the coefficients originating from the IIA metric, dilaton, and $p$-forms, respectively.

In this paper we choose yet a different solution of the section constraint. First, we impose the
existence of a Killing vector field in the IIB theory and accordingly split the coordinates $\{y^a\}=\{y^i, y^*\}$, $(i=1, \dots, 4)$, 
such that $\partial_* \Phi=0$ for all fields of the theory. Next, we relax the IIB solution,
by allowing fields to depend on the 5 coordinates
\bea
\{y^i, \tilde{y}_{*+}\}\,, \quad i=1, \dots, 4\;,
\label{yyy}
\eea
such that the section condition (\ref{sectioncondition}) is still satisfied.
In the following, we will evaluate ExFT in the IIB parametrization (\ref{M2B}) however imposing a particular
additional $\tilde{y}_{*+}$-dependence according to a simple Scherk-Schwarz ansatz
which will trigger the generalized IIB equations.
It is important to note that
the choice of coordinates (\ref{yyy}) is equivalent (after rotation of the 27 coordinates) 
to selecting the IIA coordinates $\tilde{y}_{a+}$ in (\ref{breakIIB1}).
Applying the same rotation to the IIA parametrization of ExFT fields such as (\ref{VIIA})
we would simply recover the IIA theory. This is a manifestation of the fact that the generalized IIB supergravity equations
can be obtained via T-duality from a sector of IIA supergravity. Since the framework of exceptional field theory
is manifestly duality covariant, we can simply absorb the effect of this duality
into a rotation of the extended coordinates. We will thus evaluate exceptional field theory in its IIB parametrization
(\ref{M2B}) however in coordinates (\ref{yyy}) and with a proper Scherk-Schwarz ansatz in $\tilde{y}_{*+}$
in order to obtain directly the generalized IIB equations.

\subsection{Scherk-Schwarz ansatz}

Following the previous discussion, and having chosen physical coordinates according to (\ref{yyy}) 
we now impose on the ExFT fields (\ref{fieldcontent}) 
a specific $\tilde{y}_{*+}$-dependence, such that in particular the total $\tilde{y}_{*+}$-dependence consistently
factors out from all the equations of motion. This is achieved by a Scherk-Schwarz ansatz \cite{Hohm:2014qga}
\bea\label{SSembedding}
{\cal M}_{MN} &=& U_{M}{}^{{K}}(\tilde{y})\,U_{N}{}^{{L}}(\tilde{y})\,{M}_{{K}{L}}(x^\mu,y^i)\;, 
\nonumber\\
 g_{\mu\nu} &=& \rho^{-2}(\tilde{y})\,{\rm g}_{\mu\nu}(x^\mu,y^i)\;,\nonumber\\
  {\cal A}_{\mu}{}^{M} &=& \rho^{-1}(\tilde{y})\, A_{\mu}{}^{{N}}(x^\mu,y^i)\,(U^{-1})_{{N}}{}^{M}(\tilde{y}) \;, 
  \nonumber\\
  {\cal B}_{\mu\nu\,M} &=& \,\rho^{-2}(\tilde{y})\, U_M{}^{{N}}(\tilde{y})\,B_{\mu\nu\,{N}}(x^\mu,y^i)
  \;,
 \eea
where the $\tilde{y}_{*+}$-dependence of all fields
is carried by an ${\rm E}_{6(6)}$-valued twist matrix $U_{N}{}^{{L}}$ and a scalar factor $\rho$\,.
For simplicity of the notation, here and in the following we also use the notation $\tilde{y}\equiv\tilde{y}_{*+}$\,.\footnote{
Note that the ansatz (\ref{SSembedding}) is slightly more general than the ones studied in \cite{Hohm:2014qga}
in that the fields multiplying the twist matrices on the r.h.s.\ do not only depend on the external coordinates $x^\mu$
but also on part of the internal coordinates ${y^i}$. In this sense, the ansatz (\ref{SSembedding}) 
resembles the embedding of deformations of ExFT
studied in \cite{Ciceri:2016dmd}
(and in \cite{Grana:2012rr} in the context of double field theory), 
although here all fields and twist matrices respect the section constraint,
so we remain within the original framework.
}

The relevant Scherk-Schwarz ansatz for generalized IIB supergravity is based on a twist matrix $U_M{}^N$
living in an
\bea
{\rm GL}(1) \subset {\rm SL}(2)_{\rm diag} \subset {\rm SL}(2) \times {\rm SL}(2) 
\subset {\rm SL}(2) \times {\rm SL}(6) \subset {\rm E}_{6(6)}
\;,
\eea
subgroup of the full duality group ${\rm E}_{6(6)}$\,. More precisely, upon decomposing 
\bea
{\rm E}_{6(6)}&\rightarrow& {\rm SL}(2) \times {\rm SL}(6)\;,\nonumber\\
{\bf 27} &\longrightarrow& (1,15)+(2,6')\;,
\qquad \left\{ Y^M \right\} ~\longrightarrow~ \left\{ Y^{\hat{a}\hat{b}} , \tilde{Y}_{\hat{a}\alpha} \right\}
\;,
\label{sl6}
\eea
an $({\rm SL}(2) \times {\rm SL}(6))$-valued matrix $U$ takes the form
\bea
U_M{}^{{N}} &=& 
\begin{pmatrix}
U_{\hat a}{}^{{[\hat c}} U_{\hat b{}}{}^{{\hat d}]} & 0\\
0 & (U^{-1})_{{\hat a}}{}^{\hat c} U_\alpha{}^{{\beta}}
\end{pmatrix}
\;,
\label{TU}
\eea
and we choose the matrix factors as
\bea
U_\alpha{}^{{\beta}} &=& 
\begin{pmatrix} U_+{}^+ & 0\\
0& U_-{}^- 
\end{pmatrix}
~=~
\begin{pmatrix} \rho(\tilde{y}) & 0\\
0& \rho^{-1}(\tilde{y}) 
\end{pmatrix}
\;,
\nonumber\\
U_{\hat a}{}^{{\hat b}}&=& 
\begin{pmatrix} U_i{}^j & 0&0\\
0& U_*{}^* & 0\\
0 & 0 & U_0{}^0 
\end{pmatrix}
~=~
\begin{pmatrix} \delta_i{}^j & 0&0\\
0& \rho(\tilde{y}) & 0\\
0 & 0 & \rho^{-1}(\tilde{y}) 
\end{pmatrix}
\;,
\label{UU}
\eea
with scale factor given by a linear function $\rho(\tilde{y})=\tilde{y}+c$\,.
In order to check the effect of the Scherk-Schwarz ansatz (\ref{SSembedding}) with (\ref{UU})
on the field equations of exceptional field theory, we consider the 
current
\bea
({\cal X}_{{M}}){}_{{N}}{}^{{K}} &\equiv&
\rho^{-1}\,(U^{-1})_{{M}}{}^{P} \,
(U^{-1})_{{N}}{}^{Q} \,
\partial_P U_Q{}^{{K}}
\;,
\label{ssc}
\eea
which encodes the combinations of the twist matrix and its derivatives that explicitly enter the field equations.
With the explicit form of (\ref{UU}), this current lives in the algebra $\mathfrak{sl}(2)\oplus\mathfrak{sl}(6)$ with its only
non-vanishing components given by
\bea
({\cal X}^{*+}){}_\alpha{}^{{\beta}} &=& 
\begin{pmatrix} 1 & 0\\
0& -1 
\end{pmatrix}
\;,\qquad
({\cal X}^{*+}){}_{\hat a}{}^{{\hat b}}~=~ 
\begin{pmatrix} 0_{4\times 4}& 0&0\\
0& 1& 0\\
0 & 0 & -1 
\end{pmatrix}
\;,
\label{Xi}
\eea
all constant, ensuring that the $\tilde{y}$-dependence factors out from all equations of motion.\footnote{
Strictly speaking, for consistency of the Scherk-Schwarz ansatz a weaker condition is sufficient: only the
projection of (\ref{ssc}) onto the ${\bf 27}\oplus{\bf 351}$ representation of ${\rm E}_{6(6)}$ appears in the field equations
and is required to be constant. With (\ref{Xi}) this is automatically guaranteed.
}
We have thus presented a consistent Scherk-Schwarz ansatz on the ExFT fields
which moreover satisfies the section condition. Upon explicitly evaluating the field equations,
the non-trivial $\tilde{y}$ dependence of the twist matrix induces a deformation of the
original IB equations of motion. We shall work this out in the next section.

\subsection{Induced deformation}

In this section we will illustrate with several examples how the Scherk-Schwarz ansatz (\ref{SSembedding})
induces a deformation of the resulting field equations which precisely coincides with
the deformation of the IIB field equations and Bianchi identities
discussed in section~\ref{sec:genIIB} above.
Covariant derivatives in ExFT carry 
vector fields ${\cal A}_\mu{}^M$ and internal derivatives $\partial_M$\,. 
Under $({\rm SL}(2) \times {\rm SL}(6))$, the coordinates
(\ref{yyy}) are embedded in the $Y^M$ as $\{Y^{i0}, \tilde{Y}_{*+}\}$, c.f.~(\ref{sl6}). 
With the ansatz (\ref{SSembedding}),
the relevant couplings then are obtained from
\bea
{\cal A}_\mu{}^{i0} \partial_{i0} &=& \rho^{-1}\rho\,{A}_\mu{}^{i0} \partial_{i0}
\;,\qquad
{\cal A}_\mu{}_{*+} \partial^{*+} ~=~ \rho^{-1}\rho^2\,{A}_\mu{}_{*+} \partial^{*+}
\;.
\label{ApAp}
\eea
Both operators give rise to additional $\tilde{y}$-independent couplings.
Let us e.g.\ consider the covariant derivative on the external metric (\ref{Dgg}).
With the Scherk-Schwarz ansatz (\ref{SSembedding}), we obtain via (\ref{ApAp})
\bea
 D_\mu g_{\nu\rho} &=& \rho^{-2}(\tilde{y}) \left(
 \partial_\mu {\rm g}_{\nu\rho} 
 - 2\,{ A}_\mu{}^{i0} \partial_{i0} {\rm g}_{\nu\rho}
   -\frac43\,\partial_{0i} {A}_\mu{}^{0i}\,{\rm g}_{\nu\rho}
 +\frac43\,  {A}_\mu{}_{*+}  {\rm g}_{\nu\rho} \right)
\;.
\eea
The first three terms on the r.h.s.\ correspond to the standard ExFT result and upon translation 
into the IIB fields contribute to the standard IIB field equations~\cite{Baguet:2015xha}. 
We will thus employ the notation
\bea
D_\mu g_{\nu\rho} &=& \rho^{-2}(\tilde{y}) \left( \mathring{D}_\mu {\rm g}_{\nu\rho}
 +\frac43\,  { A}_\mu{}_{*+}  {\rm g}_{\nu\rho} \right)
\;.
\label{Dg}
\eea
The last term
captures the effect of the Scherk-Schwarz twist matrix and shows that the IIB space-time metric acquires
non-trivial covariant derivatives which is precisely in accordance with our discussion above regarding
the charged IIB metric (\ref{metricTromb}) after transition to the Einstein frame.\footnote{
To be precise, after identification ${A}_\mu{}_{*+}={B}_{\mu*\,+}$, the factor 4/3 in (\ref{Dg}) comes via the
standard $5+5$ Kaluza-Klein decomposition
\bea
G_{mn} &=& \begin{pmatrix}
({\rm det} \, g_{ab})^{-1/3}\,g_{\mu\nu} + \dots & A_\mu{}^b g_{ab}\\
g_{ab}A_\mu{}^b & g_{ab} 
\end{pmatrix}
\;,
\eea
of the IIB metric.
}
The Riemann tensor whose contraction appears in the Einstein field equations will thus
correspond to the curvature of the modified connection~(\ref{metricTromb}) as in the
generalized IIB equations.

In a similar way, we can work out the ExFT field strengths (\ref{FYM}) under the Scherk-Schwarz 
ansatz (\ref{SSembedding}). As a general feature of the Scherk-Schwarz ansatz with consistent twist
matrices, the $\tilde{y}$-dependence of these field strengths consistently factors out according to
\bea
{\cal F}_{\mu\nu}{}^M(x,Y)&=&
\rho^{-1}(\tilde{y})(U^{-1})_{{N}}{}^M(\tilde{y})\,{\cal F}_{\mu\nu}{}^M(x,y)
\;,
\eea
where
\bea
{\cal F}_{\mu\nu}{}^M(x,y)&\equiv& 
 \mathring{\cF}_{\mu\nu}{}^{{NM}}
 +X_{{K}{L}}{}^{{M}} \left(A_{[\mu}{}^{{K}} A_{\nu]}{}^{{L}} -2 \,d^{KLN} \, B_{\mu\nu\, {N}}\right)
 \;,
 \label{FXB}
\eea
describes a deformation of the standard ExFT field strength $\mathring{\cF}_{\mu\nu}{}^{{M}}$ by non-abelian
terms carrying the generic structure of five-dimensional gauged supergravity~\cite{deWit:2004nw}
encoded in an embedding tensor $X_{MN}{}^K$ living in the ${\bf 351}+{\bf 27}$ representation of E$_{6(6)}$\,.
Within the Scherk-Schwarz ansatz, the embedding tensor is obtained from projecting 
(\ref{ssc}) onto the relevant E$_{6(6)}$ representations.
Again, the form of (\ref{FXB}) resembles the deformations of ExFT studied in \cite{Ciceri:2016dmd}, 
although here it simply results from a Scherk-Schwarz ansatz within the original ExFT.
Structurewise, the new couplings (\ref{FXB}) resemble those introduced in (\ref{FFF}) in order
to account for the deformed Bianchi identities in generalized IIB supergravity. In the rest of this paper,
we will make the agreement precise using the explicit dictionary between ExFT and IIB 
fields~\cite{Baguet:2015xha}.

Working out  (\ref{FYM}), it follows that
the twist matrix (\ref{TU})--(\ref{UU}) induces an embedding tensor
\bea
X_{MN}{}^K &=& (\tilde{X}_{M})_N{}^K +\frac23\,\delta_M{}^{*+}\,\delta_N{}^K
\;,
\label{XX}
\eea
in (\ref{FXB}). 
Upon contraction with a gauge parameter $\Lambda^M$ it identifies the 
gauged generators within $\mathfrak{e}_{6(6)} \oplus \mathbb{R}_{\rm tromb}$.
The second term in (\ref{XX})
refers to the gauging of the trombone symmetry under which the ExFT fields
$\left\{g_{\mu\nu}, {\cal M}_{MN},{\cal A}_\mu{}^M, {\cal B}_{\mu\nu\,M} \right\}$ 
scale with weight $\{2,0,1,2\}$, respectively,
whose effect we have already observed in (\ref{Dg}). The first term in (\ref{XX})
identifies the gauged generators within $\mathfrak{e}_{6(6)}$,
combining the diagonal generators
\bea
\left(\Lambda^{M} \tilde X_{M}\right)_{\mathfrak{sl}(2)}
=
\begin{pmatrix}
 \frac12\,\Lambda_{*+}&0\\
- \Lambda_{*-}&  -\frac12\,\Lambda_{*+}
\end{pmatrix}\;,
\quad
\left(\Lambda^{M} \tilde X_{M}\right)_{\mathfrak{sl}(6)}
=
\begin{pmatrix}
\frac1{6} \, \Lambda_{*+}\,\mathbb{I}_4 &0&0\\
0&\frac1{6} \, \Lambda_{*+} & 0\\
0&  \Lambda_{0+} &-\frac{5}{6} \, \Lambda_{*+}
\end{pmatrix}
\;,\;\;
\label{X_slsl}
\eea
within $\mathfrak{sl}(2)\oplus\mathfrak{sl}(6)$
with the
off-diagonal generators
\bea
(\Lambda^{M}\tilde X_{M}){}^{*+,ij} &=&(\Lambda^{M}\,\tilde X_{M}){}^{i+,j*} ~=~
-\Lambda^{ij}
\;,
\nonumber\\
(\Lambda^{M}\tilde X_{M})_{ij,0-} &=& (\Lambda^{M}\,\tilde X_{M})_{0i,j-} ~=~
-\frac12\,\varepsilon_{*ijkl} \, \Lambda^{kl}\,
\;,
\label{X_off}
\eea
 in $\mathfrak{e}_{6(6)}\backslash\, (\mathfrak{sl}(6)\oplus\mathfrak{sl}(2))$.
 The St\"uckelberg-type couplings in (\ref{FXB}) to the two-forms $B_{\mu\nu\,M}$
 are read off from (\ref{X_slsl}), (\ref{X_off}) together with the explicit form of $d^{MNP}$
 in the decomposition (\ref{breakIIB1}), see \cite{Baguet:2015xha}.
The explicit result for the various components of the field strengths (\ref{FXB}) is the following
\bea\label{FieldSTrengths}
{\cal F}_{\mu\nu\,a+} &=&\mathring{\cal F}_{\mu\nu\,i+}
\;,
\nonumber\\
{\cal F}_{\mu\nu\,a-} &=&\mathring{\cal F}_{\mu\nu\,a-}
+ A_{[\mu\,*+}{} A_{\nu]\,a-}
-A_{[\mu\,*-}{} A_{\nu]\,a+}
+\sqrt{2}\,\tilde{B}_{\mu\nu\,a*}
\;,
\nonumber\\
{\cal F}_{\mu\nu\,abc} &=&\mathring{\cal F}_{\mu\nu\,abc}
+2\,A_{[\mu}{}_{*+} A_{\nu]}{}_{abc} 
+\frac3{2\sqrt{2}}\,\varepsilon_{abcd*}\, \tilde B_{\mu\nu}{}^{d-}
\;,
\nonumber\\
{\cal F}_{\mu\nu\,+} &=& \mathring{\cal F}_{\mu\nu\,+} 
+2\, A_{[\mu\,*+}{} A_{\nu]\,+}
\;,
\label{FXF}
\eea
with the redefined  two-forms
\bea
\tilde {B}_{\mu\nu\,ab} &\equiv& \sqrt{10}\, { B}_{\mu\nu\,ab} + {A}_{[\mu}{}^{c}  {A}_{\nu]\,abc}
\;,\qquad
\nonumber\\
\tilde{B}_{\mu\nu}{}^{k\alpha} &\equiv& 
\sqrt{10}\, {B}_{\mu\nu}{}^{a\alpha}+\varepsilon^{\alpha\beta}\,{A}_{[\mu}{}^a  {A}_{\nu]}{}_\beta
+\frac{\sqrt{2}}{6}\,\varepsilon^{\alpha\beta}\,\varepsilon^{abcde}\,A_{[\mu}{}_{|b\beta|} A_{\nu]}{}_{cde}
\;.
\eea
Comparing the deformed field strengths (\ref{FXF}) to the field strengths (\ref{FFF}) 
solving the Bianchi identities of generalized IIB supergravity, we find precise agreement
upon identifying the ExFT components with the IIB field strengths
(the precise dictionary between fields has been given in \cite{Baguet:2015xha}
and in particular takes care of the $\sqrt{2}$ factors that arise in the ExFT expressions (\ref{FXF})).

Of course, the field strengths (\ref{FXF}) only represent part of the full IIB field strengths,
in which two of the ten-dimensional indices are chosen to be external. The remaining IIB components
will appear among other ExFT fields. E.g.\ let us consider the three-form field strength
${\cal H}_{\mu\nu\rho\,M}$ defined by (\ref{defH}). Evaluating this definition with the above Scherk-Schwarz
ansatz in particular yields the components
\bea
{\cH}_{\mu\nu\rho\,-}&=&\mathring{{\cH}}_{\mu\nu\rho\,-}
+\sqrt{2}\,{\cal O}_{\mu\nu\rho}
\;,
\nonumber\\
{\cH}_{\mu\nu\rho\,*a}&=&\mathring{\tilde{\cH}}_{\mu\nu\rho\,*i}+3\,A_{[\mu|\,*+|}\tilde{B}_{\nu\rho] \, *a}
+\frac{3\sqrt{2}}{2}\,A_{[\mu| \, *+|}A_{\nu|\,*-|}A_{\rho] \,a+}-\partial_a\,{\cal O}_{\mu\nu\rho}
\;.
\label{HH}
\eea
The second and third term of ${\cH}_{\mu\nu\rho\,*i}$ reproduce the 
corresponding deformation terms in (\ref{FFF}).
The term ${\cal O}_{\mu\nu\rho}$ in (\ref{HH}) 
denotes the undetermined contribution in the field strength 
which vanishes under the projection $d^{KMN}\partial_N$ in (\ref{defH}).
In the undeformed IIB theory, this term is already present in ${\cH}_{\mu\nu\rho\,*i}$\,.
It arises as an integration constant in the ExFT field equations and is
identified with a component of the IIB four-form according to
\bea
\sqrt{2}\,{\cal O}_{\mu\nu\rho}&=&
C_{\mu\nu\rho*}+
\frac32\,B_{[\mu|*+|} B_{\nu\rho]\,-} - \frac32\,B_{[\mu|*-|} B_{\nu\rho]\,+} 
\;,
\eea
in order to reconstruct the selfdual IIB five-form field strength from ExFT.
In the deformed case we are considering here, the same ${\cal O}_{\mu\nu\rho}$
arises as part of ${\cH}_{\mu\nu\rho\,-}$ in (\ref{HH}) where it precisely acounts for the
deformation of the IIB three-form field strength ${F}_{\mu\nu\rho}$, see (\ref{FFF}).
Again we thus find complete agreement.

In a similar way, the deformed scalar currents ${\cal M}^{MK} D_\mu {\cal M}_{KN}$ 
with the block decomposition (\ref{M2B}) and parametrization (\ref{V27AB}) can be
matched to the corresponding components of (\ref{FFF}) in which one 
of the ten-dimensional indices  is chosen to be external.
Thus all the building blocks of the ExFT Lagrangian (\ref{EFTaction}) exhibit precisely
the deformations of their IIB counterparts (\ref{FFF}).
Since equations (\ref{FFF}) were derived as solution of the deformed IIB Bianchi identities,
it follows that after imposing the Scherk-Schwarz ansatz (\ref{SSembedding}), the ExFT fields
satisfy the deformed IIB Bianchi identities. Moreover, most of the generalized IIB field equations
are obtained by covariantization of the standard IIB equations, i.e.\ by replacing the IIB field
strengths by their deformed expressions (\ref{FFF}). This is true for the Einstein field equations
(upon taking into account the charged metric in the Einstein frame, c.f.\ (\ref{metricTromb})) and
the self-duality equation ${\cal F}_{rmnpq}=\star{\cal F}_{rmnpq}$ for the five-form field
strength. Upon using the explicit dictionary between ExFT fields and IIB fields \cite{Baguet:2015xha}
these equations thus follow from the ExFT dynamics after imposing the Scherk-Schwarz ansatz. 
The two-form field equations (\ref{DFJ}) in generalized IIB supergravity on the other hand
are not only covariantized via (\ref{FFF}) but also acquire a source term $J^{mn}{}_\beta$. 
In ExFT, the analogous term descends from
variation of the Lagrangian (\ref{EFTaction}) w.r.t.\ the gauge fields which upon implementing the
Scherk-Schwarz ansatz gives rise to additional source terms from the Einstein-Hilbert term and the
scalar kinetic term.

\section{Conclusions}

In this paper, we have shown how the equations of generalized IIB supergravity found in \cite{Arutyunov:2015mqj}
can naturally be obtained from exceptional field theory upon imposing a simple Scherk-Schwarz type ansatz on 
all the fields that captures their non-isometric behavior in the IIA theory. 
The Scherk-Schwarz ansatz satisfies the consistency equations \cite{Hohm:2014qga}
and moreover the section constraints (\ref{sectioncondition}) and induces a deformation of the standard 
IIB supergravity equations.
We have verified explicitly for most of their components that the deformed ExFT fields coincide with the
deformed IIB field strengths (\ref{FFF}) which have been determined by solving the deformed IIB
Bianchi identities.
The Scherk-Schwarz ansatz straightforwardly extends to the 
fermionic extension of ExFT~\cite{Musaev:2014lna}.
It should thus also be possible to reproduce from exceptional field theory
the fermionic completion of the generalized IIB equations
that has been worked out in~\cite{Wulff:2016tju}.
We should stress that although exceptional field theory admits a Lagrangian formulation
(\ref{EFTaction}) this does not allow to conclude the existence of an action underlying the generalized IIB equations,
since the Scherk-Schwarz ansatz (\ref{SSembedding}) is imposed on the level of the field
equations and not on the action. The appearance of a trombone gauging (\ref{XX}) in the ExFT
formulation is in fact a sign that the resulting field equations cannot be obtained from an action~\cite{LeDiffon:2008sh}.

We have in this article embedded the generalized IIB equations 
into E$_{6(6)}$ ExFT. In principle, the same construction can be
repeated for any of the exceptional field theories upon an appropriate
decomposition of the IIB coordinates into external and internal 
coordinates. If one is merely interested in reproducing the generalized IIB equations
from a covariant framework, the most efficient framework is the 
SL(2) ExFT from \cite{Berman:2015rcc}. In this case, 9-dimensional
covariance is kept manifest throughout the construction and other than
splitting off the 10th coordinate $y^*$, we avoid the split
into internal $\{y^i\}$ and external $\{x^\mu\}$ coordinates.
Here, we have given the embedding into E$_{6(6)}$ ExFT since it is 
this formulation which most naturally carries the $AdS_5 \times S^5$ background
in the undeformed case.
In particular, in E$_{6(6)}$ ExFT it follows from a Scherk-Schwarz reduction ansatz 
that the fluctuations 
around the $AdS_5 \times S^5$ background give rise to a consistent truncation
of IIB supergravity to a five-dimensional maximal supergravity theory~\cite{Hohm:2014qga,Baguet:2015sma}.
It will be very interesting to check if similarly the ABF background allows for
a compact formulation within E$_{6(6)}$ ExFT, combining the Scherk-Schwarz
ansatz (\ref{UU}) with a deformation of the Scherk-Schwarz twist underlying the $S^5$ geometry
and an $x^\mu$-dependent solution of a five-dimensional supergravity. 
In particular, this would allow to address the question of consistent truncations around the ABF background
and be of particular interest in view of possible holographic interpretations
of the deformation. We hope to come back to these questions in the near future.

\subsection*{Acknowlegements}
We would like to thank Riccardo Borsato, Olaf Hohm and Mario Trigiante for helpful discussions.
H.S.\ thanks the Galileo Galilei Institute for Theoretical Physics for the hospitality and the
INFN for partial support during the completion of this work.
This work is partially supported by the French Agence Nationale de la Recherche (ANR) 
under grant ANR-15-CE31-0006 DefIS.


\providecommand{\href}[2]{#2}\begingroup\raggedright\endgroup

\end{document}